# A Multiscaling Fingerprint of Earthquake Diffusion in Seismic Swarms

**Cataldo Godano**[1] **and Giuseppe Petrillo**[2]

[1]Dipartimento di Matematica e Fisica - Università della Campania
[2]CNRS, Laboratoire de Physique, ENS de Lyon, Univ. Lyon, France

**Key Points:**

- Earthquake swarms exhibit a characteristic multiscaling diffusion signature.
- Small and large interevent distances evolve with different effective scaling laws.
- Multiscaling distinguishes seismic swarms from conventional tectonic sequences.

Corresponding author: Cataldo Godano, cataldo.godano@unicampania.it

**Abstract**
Seismic swarms are commonly associated with fluid migration and other transient processes, yet their spatial migration remains difficult to quantify using conventional diffusion models. Here we analyze ten persistent earthquake swarms identified within the relocated Southern California earthquake catalogue using a consensus clustering approach. We characterize their migration through the multiscaling spectrum $\varepsilon(q)$ obtained from the temporal evolution of the moments of interevent distances. All analyzed swarms exhibit a common scaling signature: the spectrum is approximately linear for negative moments but departs systematically from a single linear behavior for positive moments, indicating that small and large interevent distances evolve with different effective scaling laws. In contrast, the Landers tectonic sequence displays a high-order saturation of $\varepsilon(q)$, consistent with spatially bounded diffusion. These results reveal that earthquake swarms are characterized by strong anomalous diffusion and suggest that the multiscaling spectrum provides a quantitative fingerprint capable of distinguishing swarm migration from conventional tectonic earthquake sequences.

## Plain Language Summary

Earthquakes often occur in clusters known as seismic swarms, where many events take place within a limited area without a single dominant mainshock. Understanding how these swarms migrate is important because their evolution may reflect fluid migration, volcanic activity, or slow fault motion. We analyzed ten large swarms in Southern California and found that short and long earthquake separations evolve differently through time. This multiscale behavior distinguishes swarms from a classical tectonic earthquake sequence and provides a new statistical fingerprint of earthquake migration.

## 1 Introduction

Seismic swarms represent one of the most intriguing forms of clustered seismicity because they develop without a dominant mainshock and are commonly associated with transient physical processes such as fluid migration, volcanic unrest, or aseismic fault slip. Originally defined by Mogi (1963) as sequences of multiple earthquakes occurring within a localized area over a relatively short period of time, swarms differ from classical earthquake sequences in that they consist of numerous events of comparable magnitude rather than a single large mainshock followed by aftershocks (Mogi, 1963; Vidale & Shearer, 2006). Although this definition was proposed several decades ago, it still captures the essential characteristics of swarm activity despite the dramatic improvements in earthquake monitoring.

Earthquakes rarely occur at a constant rate, but instead cluster in bursts or episodes of enhanced activity (Hainzl, 2004; Vidale & Shearer, 2006). Seismic swarms are typically confined to relatively small spatial regions, frequently associated with active fault systems, geothermal areas, or volcanic structures (Hill, 1977; Ross et al., 2025). Unlike ordinary tectonic sequences, which are primarily driven by coseismic stress redistribution, swarms are generally interpreted as the response of fault systems to transient forcing mechanisms. The most widely accepted triggering processes include pore-pressure diffusion caused by underground fluids (Hubbert & Rubey, 1959; Ellsworth, 2013), magma or magmatic-fluid migration in volcanic environments (Tramelli et al., 2021; Convertito et al., 2025), and slow aseismic fault slip loading small locked asperities (Nadeau & McEvilly, 1999; Sammis & Rice, 2001; Uchida & Beroza, 2024).

The temporal evolution of seismic swarms often differs substantially from that of ordinary aftershock sequences. Episodes of rapidly accelerating seismicity have been reported in several volcanic and geothermal regions, including the 2000 Izu Islands swarm (Toda et al., 2002). As discussed by Llenos and Michael (2019), the Omori law alone is frequently insufficient to describe swarm activity. This behavior has been interpreted as evidence for a nonstationary background seismicity rate (Petrillo et al., 2024), while alternative analytical descriptions based on Gamma-like occurrence rates have also been proposed (Godano et al., 2023).

Besides their temporal evolution, seismic swarms exhibit complex spatial migration patterns whose physical origin remains incompletely understood. Earthquake diffusion was first investigated by Reasenberg and Simpson (1992) for the Loma Prieta sequence and has subsequently been documented in both tectonic and volcanic environments (Llenos et al., 2009; Florez et al., 2025). Characterizing this migration is fundamental for discriminating between different physical mechanisms, including tectonic stress transfer, viscoelastic relaxation, rate-and-state friction, pore-pressure diffusion, and magma intrusion. In homogeneous porous media, fluid-driven seismicity is commonly modeled as a classical diffusion process in which the triggering front expands as

$$r \sim t^{1/2}.$$

Natural fault systems, however, are highly heterogeneous, containing complex fracture networks, anisotropic permeability structures, and geometrical barriers that can substantially modify transport properties. Consequently, earthquake migration may depart from classical diffusion and exhibit anomalous scaling behavior.

A natural framework for investigating these heterogeneous transport processes is provided by the multiscaling formalism introduced by Godano and Pingue (2005). Rather than describing earthquake migration through a single diffusion exponent, multiscaling examines how different statistical moments of the interevent-distance distribution evolve with time. A linear scaling spectrum indicates that all spatial scales evolve according to a common diffusion law, whereas nonlinear spectra reveal the coexistence of multiple effective scaling behaviors, a hallmark of strong anomalous diffusion (Castiglione et al., 1999; Paladin & Vulpiani, 1987). Despite its potential, this framework has so far been applied only to general earthquake catalogues, and its ability to characterize seismic swarms has not yet been systematically investigated.

Here we address this question by analyzing persistent seismic swarms identified within the relocated Southern California earthquake catalogue. Swarms are first identified using a consensus clustering procedure that combines multiple space-time clustering realizations. We then investigate the multiscaling properties of earthquake migration through the scaling spectrum of the interevent-distance moments and compare these results with a classical tectonic sequence. We show that persistent seismic swarms exhibit a remarkably consistent nonlinear scaling spectrum, whereas the Landers earthquake sequence displays high-order saturation. These contrasting behaviors suggest that the multiscaling spectrum provides a quantitative statistical fingerprint capable of distinguishing swarm migration from conventional tectonic earthquake sequences.

## 2 Methods and Data

We analyze the relocated Southern California earthquake catalogue of Hauksson et al. (2012), downloaded from the Southern California Earthquake Data Center (SCEDC). Before identifying earthquake swarms, we estimate the catalogue completeness magnitude using the coefficient of variation (CV) method (Godano & Petrillo, 2023; Godano, Petrillo, & Lippiello, 2024). The coefficient of variation is defined as

$$c_V = \frac{\sigma}{\mu}, \tag{1}$$

where $\sigma$ and $\mu$ denote the standard deviation and the mean of the earthquake magnitudes above a given threshold magnitude $m_{th}$. For an exponential distribution, corresponding to the Gutenberg-Richter law, the coefficient of variation is equal to one. Consequently, the completeness magnitude can be estimated by evaluating $c_V$ as a function of $m_{th}$ and identifying the threshold above which the magnitude distribution becomes compatible with an exponential law.
Following Godano, Tramelli, et al. (2024), we adopt a conservative threshold $c_V = 0.9$, which provides a more robust estimate than the value 0.73 originally proposed from numerical simulations. This analysis yields a completeness magnitude of $m_c = 2.0$. We remark that this estimate does not account for the spatial and temporal variability of completeness, which can become significant immediately after large earthquakes because of short-term aftershock incompleteness (STAI).

Earthquake swarms are then identified using a consensus clustering approach inspired by the workflow proposed by Passarelli et al. (2026). In their study, earthquake clusters are first detected using different declustering or clustering algorithms and subsequently classified as swarm-like or mainshock–aftershock sequences through the temporal distribution of seismic moment release, quantified by the centroid time ($t_c$), skewness ($S$) and kurtosis ($K$) of the normalized moment release function. Synthetic ETAS catalogues showed that swarm-like sequences are typically characterized by $t_c > 0.5, S < 10, K < 125$, where $t_c$ is the moment-weighted centroid time, $S$ the skewness and $K$ the kurtosis of the seismic moment release distribution. Rather than relying on a single clustering realization, we extend this methodology by introducing a consensus procedure that minimizes the dependence on the arbitrary choice of clustering parameters.
Clusters are detected using a DBSCAN-like algorithm based on the normalized space-time metric

$$\eta_{ij} = \left[ \left( \frac{|t_i - t_j|}{\tau} \right)^2 + \left( \frac{r_{ij}}{r_0} \right)^2 \right]^{1/2}, \tag{2}$$

where $r_{ij}$ is the epicentral distance between earthquakes, computed through the haversine formula, $\tau$ is the characteristic temporal scale and $r_0$ the characteristic spatial scale. Two earthquakes are considered neighbors whenever $\eta_{ij} \leq \epsilon$. Only clusters containing at least 30 earthquakes are retained.
To reduce the sensitivity of the results to the specific clustering parameters, the algorithm is repeated over the parameter grid

$$r_0 = (5,\ 10,\ 15)\ \text{km},$$

$$\tau = (0.03,\ 0.05,\ 0.10)\ \text{yr},$$

$$\epsilon = (0.8,\ 1.0,\ 1.2),$$

leading to 27 independent clustering realizations. For every detected cluster, earthquake magnitudes are converted into seismic moment according to

$$M_0 = 10^{1.5M+9.1}, \tag{3}$$

and the centroid time, skewness and kurtosis of the normalized moment release are computed following Passarelli et al. (2026). Clusters satisfying the swarm criteria reported above are classified as swarm-like. Finally, we combine the information from all clustering realizations. For every earthquake $i$, we compute the membership probability

$$p_i = \frac{N_i}{N_{\text{run}}}, \tag{4}$$

where $N_i$ is the number of realizations in which event $i$ belongs to a swarm-like cluster. Similarly, for every pair of earthquakes $(i, j)$ we compute the co-membership probability

$$P_{ij} = \frac{N_{ij}}{N_{\text{run}}}, \tag{5}$$

where $N_{ij}$ denotes the number of realizations in which the two earthquakes belong to the same swarm-like cluster. Persistent swarms are finally obtained as the connected components of the graph formed by earthquakes satisfying

$$p_i \geq 0.7$$

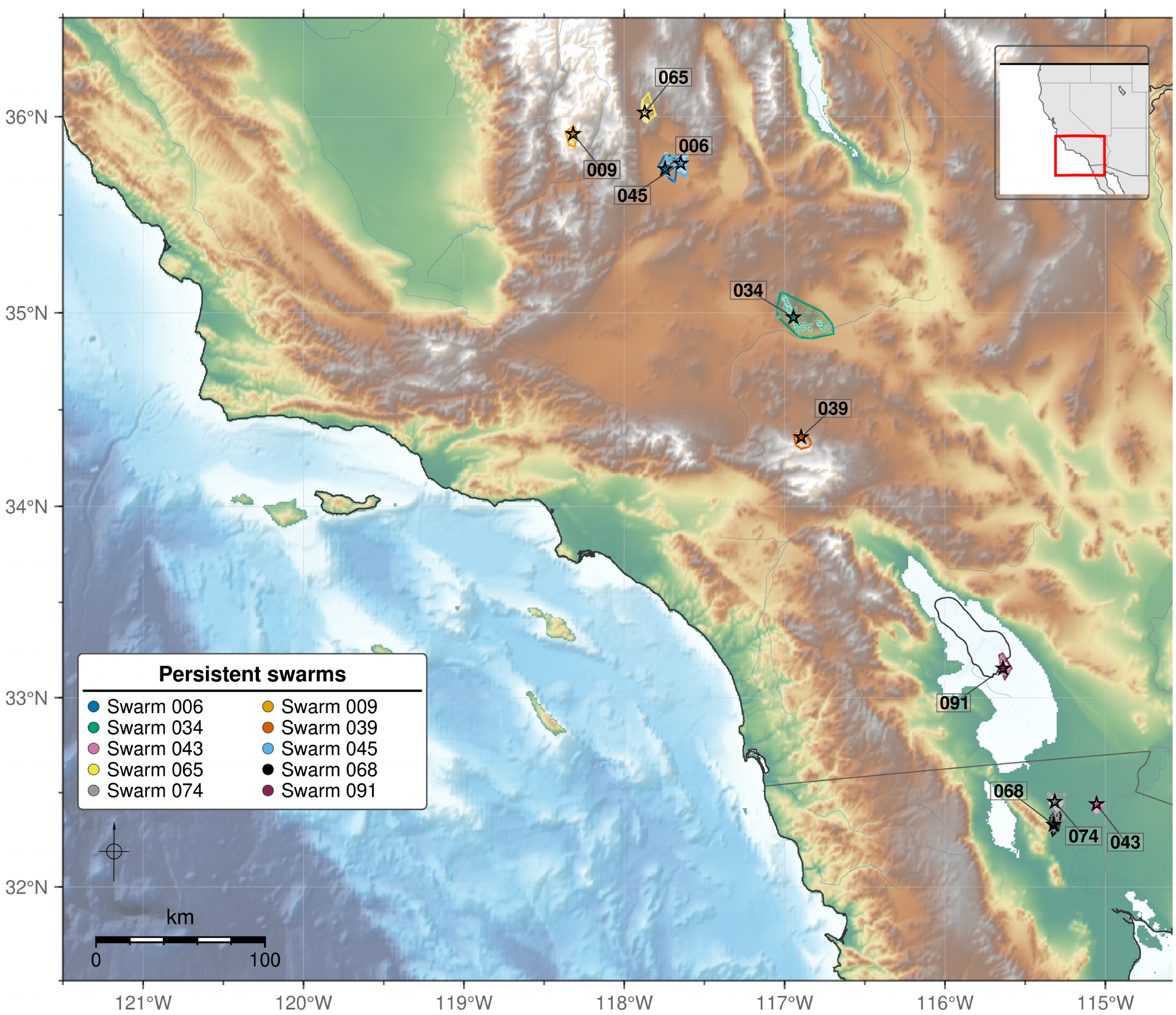

**Figure 1.** Spatial distribution of the ten persistent earthquake swarms analyzed in this study. Colored circles show earthquake epicenters, with symbol size scaled according to magnitude. Colored polygons outline the convex hull of each swarm, while stars mark their median epicentral locations. Numbers identify the swarm labels used throughout the manuscript. The inset shows the location of the study region within western North America.

and connected through links satisfying

$$P_{ij} \geq 0.7.$$

This consensus strategy substantially reduces the dependence of the final catalogue on any individual clustering realization, retaining only events and event pairs that are consistently classified as belonging to the same swarm over a broad range of spatial and temporal scales. The procedure identifies 95 persistent earthquake swarms in Southern California. Since the multiscaling analysis requires sufficiently populated catalogues to obtain stable estimates of the scaling exponents, only the 10 swarms containing more than 200 earthquakes are analyzed in the following.

Their spatial distribution is shown in Figure 1.

## 3 Multiscaling in Earthquake Diffusion

The multiscaling properties of earthquake migration are investigated through the moments of the spatial separation between earthquake pairs (Castiglione et al., 1999; Go-dano & Pingue, 2005). For each temporal lag $\theta$, we define

$$M_q(\theta) \equiv \langle r_{ij}^q \rangle_\theta = \frac{\sum_{i<j} r_{ij}^q \delta_{\Delta\theta}(\theta - \theta_{ij})}{\sum_{i<j} \delta_{\Delta\theta}(\theta - \theta_{ij})}, \tag{6}$$

where $r_{ij}$ is the epicentral distance between earthquakes $i$ and $j$, $\theta_{ij} = t_j - t_i$ is their temporal separation, and $\delta_{\Delta\theta}$ denotes a finite-width temporal bin centered at $\theta$. Thus, $M_q(\theta)$ is the average value of $r_{ij}^q$ over all earthquake pairs whose temporal separation falls within the selected bin. Within a finite range of temporal lags, the moments exhibit the scaling behavior

$$M_q(\theta) \sim \theta^{\varepsilon(q)}, \tag{7}$$

where $\varepsilon(q)$ is estimated as the slope of $\log M_q(\theta)$ versus $\log\theta$. If earthquake migration is characterized by a single diffusion exponent $H$, the typical interevent distance scales as

$$r \sim \theta^H. \tag{8}$$

Consequently, the moments defined in Equation 6 satisfy

$$M_q(\theta) = \langle r^q \rangle_\theta \sim \theta^{qH}, \tag{9}$$

and therefore

$$\varepsilon(q) = qH. \tag{10}$$

A linear dependence of $\varepsilon(q)$ on $q$ therefore indicates that all moments of the interevent-distance distribution are governed by a single scaling exponent. Conversely, a significant nonlinear dependence implies that different moments scale with different effective exponents. This behavior is commonly referred to as strong anomalous diffusion (Castiglione et al., 1999).

The scaling exponents $\varepsilon(q)$ are evaluated over the range $-2 \leq q \leq 15$. Negative-order moments emphasize the shortest interevent distances and are consequently particularly sensitive to location uncertainty and nearly coincident epicenters. We therefore do not consider values below $q = -2$. Conversely, positive-order moments increasingly weight the largest spatial separations. To reduce the sensitivity of negative-order moments to spatial separations below the reliable catalogue resolution, earthquake pairs with $r_{ij} \leq 0.05$ km are excluded. Temporal lags are grouped into logarithmic bins whose successive edges differ by a factor of 1.3.

For each swarm, a single temporal scaling interval is used for all values of $q$. Candidate intervals are required to contain at least six consecutive logarithmic bins, to span at least 0.7 decades in temporal lag, and to contain at least 50 earthquake pairs per bin. Among the admissible intervals, we select the range that provides the best overall linear scaling of $\log M_q(\theta)$ versus $\log\theta$ for a representative set of moment orders. Keeping the same temporal interval for every $q$ avoids introducing artificial curvature in $\varepsilon(q)$ by fitting different moments over different ranges of $\theta$.

A convenient theoretical framework for describing a nonlinear dependence of $\varepsilon(q)$ is provided by the multifractal formalism (Paladin & Vulpiani, 1987). In this description, the observed scaling spectrum results from the coexistence of subsets characterized by different local scaling exponents $h$. Formally,

$$\varepsilon(q) = \min_h \left[qh + d - D(h)\right], \tag{11}$$

where $D(h)$ is the fractal dimension of the subset characterized by the local exponent $h$, and $d$ is the dimension of the embedding space. Since the present analysis is based on epicentral coordinates, we take $d = 2$. If only one scaling exponent contributes, Equation 11 reduces to a linear function of $q$. A nonlinear spectrum instead indicates that different portions of the displacement statistics are associated with different effective scaling exponents. Accordingly, the multifractal formalism provides a phenomenological description of the heterogeneous transport revealed by the nonlinear behavior of $\varepsilon(q)$.

The scaling spectra $\varepsilon(q)$ obtained for the ten swarms are shown in Figure 2. A common feature of nearly all the analyzed sequences is the presence of two distinct regimes. For negative values of $q$, $\varepsilon(q)$ is approximately linear, whereas for positive values of $q$ it progressively departs from a single linear trend. Swarm 43 represents a notable exception and is discussed separately below.

By definition, $M_0(\theta) = 1$, because $r_{ij}^0 = 1$ for every earthquake pair. Consequently, $\varepsilon(0) = 0$ identically and does not provide an independent estimate of the fractal dimension of the epicentral distribution. The near-zero intercepts of the fitted spectra are therefore consistent with the mathematical normalization of the moments, rather than evidence for a space-filling epicentral geometry. The approximately linear behavior observed for $q < 0$ indicates that the small-distance portion of the pairwise-distance distribution can be described, over the investigated range, by an approximately single effective scaling exponent. Because negative-order moments emphasize the shortest interevent distances, this regime primarily reflects the evolution of the most spatially compact earthquake pairs. The negative-order branch is characterized by fitting only values with $q < 0$ according to

$$\varepsilon(q) = H_{-}q, \tag{12}$$

where the intercept is constrained to zero, consistently with the exact identity $\varepsilon(0) = 0$. The coefficient $H_{-}$ represents the effective scaling exponent associated with the shortest interevent distances. For the positive-order branch, we use values with $q \geq 1$ and compare linear, quadratic, and cubic polynomial descriptions of $\varepsilon(q)$. The preferred degree is selected using the corrected Akaike information criterion (AICc). When two models differ by less than two AICc units, the lower-degree model is retained. The selected polynomials are used as empirical descriptions of the spectra and not as mechanistic models of earthquake migration.

For positive values of $q$, the spectra depart systematically from a single linear trend. Since positive-order moments increasingly weight the largest interevent distances, this nonlinear branch indicates that large spatial separations do not evolve with the same effective scaling exponent as the most compact earthquake pairs. The swarm migration process therefore cannot be characterized by a single diffusion exponent and is consistent with strong anomalous diffusion. Within a multifractal interpretation, the nonlinear dependence of $\varepsilon(q)$ is consistent with heterogeneous transport involving a hierarchy of effective scaling behaviors. In this picture, typical and extreme spatial separations contribute differently to the observed migration statistics. Similar nonlinear moment scaling has been reported in other intermittent and heterogeneous transport processes, including fully developed turbulence and anomalous diffusion (Frisch & Parisi, 1980; Castiglione et al., 1999; Seckler et al., 2024). Although all swarms exhibit the same qualitative separation between negative- and positive-order moments, the detailed shape of the positive branch differs among sequences. Several swarms display a progressively increasing convex spectrum, whereas others exhibit flattening or changes in curvature for the largest positive moments. These differences suggest that, although the migration of all swarms is consistent with strong anomalous diffusion, the statistics of the largest spatial separations are not universal. The AICc-selected polynomial fits provide a convenient empirical description of these differences without imposing a common functional form.

At the opposite limit for large $q$ values the Landers earthquake sequence exhibits a very clear saturation of the $\varepsilon(q)$ spectrum (Fig. 3). This is a signature of caged dif-fusion: epicenters cannot diffuse over a certain scale. Such a result confirms the one already obtained by Godano and Pingue (2005) for a whole California catalogue. Caged motion has been observed in condensed matter and especially in super-cooling liquids at the glassy transition. Physically, it describes a particle temporarily trapped within a "cage" formed by neighboring particles. The particle moves rapidly on microscopic scales inside the cage (sub-diffusive or local ballistic regime) but requires a very long relaxation time to break out of the cage and migrate on macroscopic scales (Weeks & Weitz, 2002; Götze, 1999). For the Landers earthquake sequence, epicentral diffusion at small scales ($q < 0$) is driven by fluid migration (Bosl & Nur, 2002; Fialko, 2004), with $\varepsilon(q)$ following a linear trend. Conversely, at larger scales ($q > 0$), the spectrum of exponents saturates at a constant value ($\approx$0.1), revealing that the epicenters are 'caged' along a quasi-linear trend that represents the horizontal projection of the fault plane.

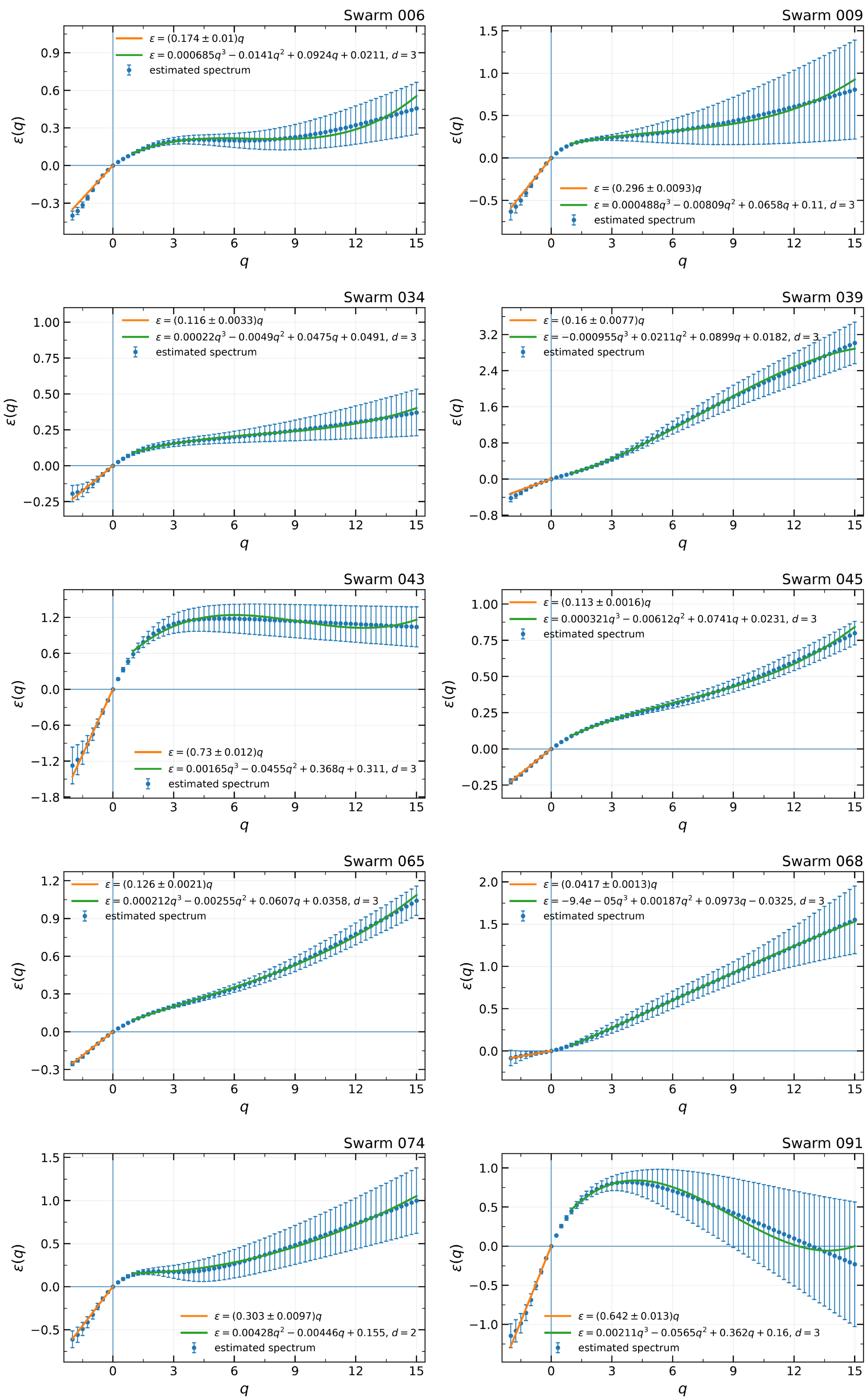

**Figure 2.** Scaling spectra $\varepsilon(q)$ for the ten persistent earthquake swarms. Symbols denote the exponents estimated from the slopes of $\log M_q(\theta)$ versus $\log \theta$, and error bars represent the corresponding regression standard errors. The negative-order branch is fitted by a straight line constrained through $\varepsilon(0) = 0$. For $q \geq 1$, the continuous curve is the lowest-degree polynomial supported by AICc among linear, quadratic, and cubic models. Each panel is shown with an independently adjusted vertical scale.

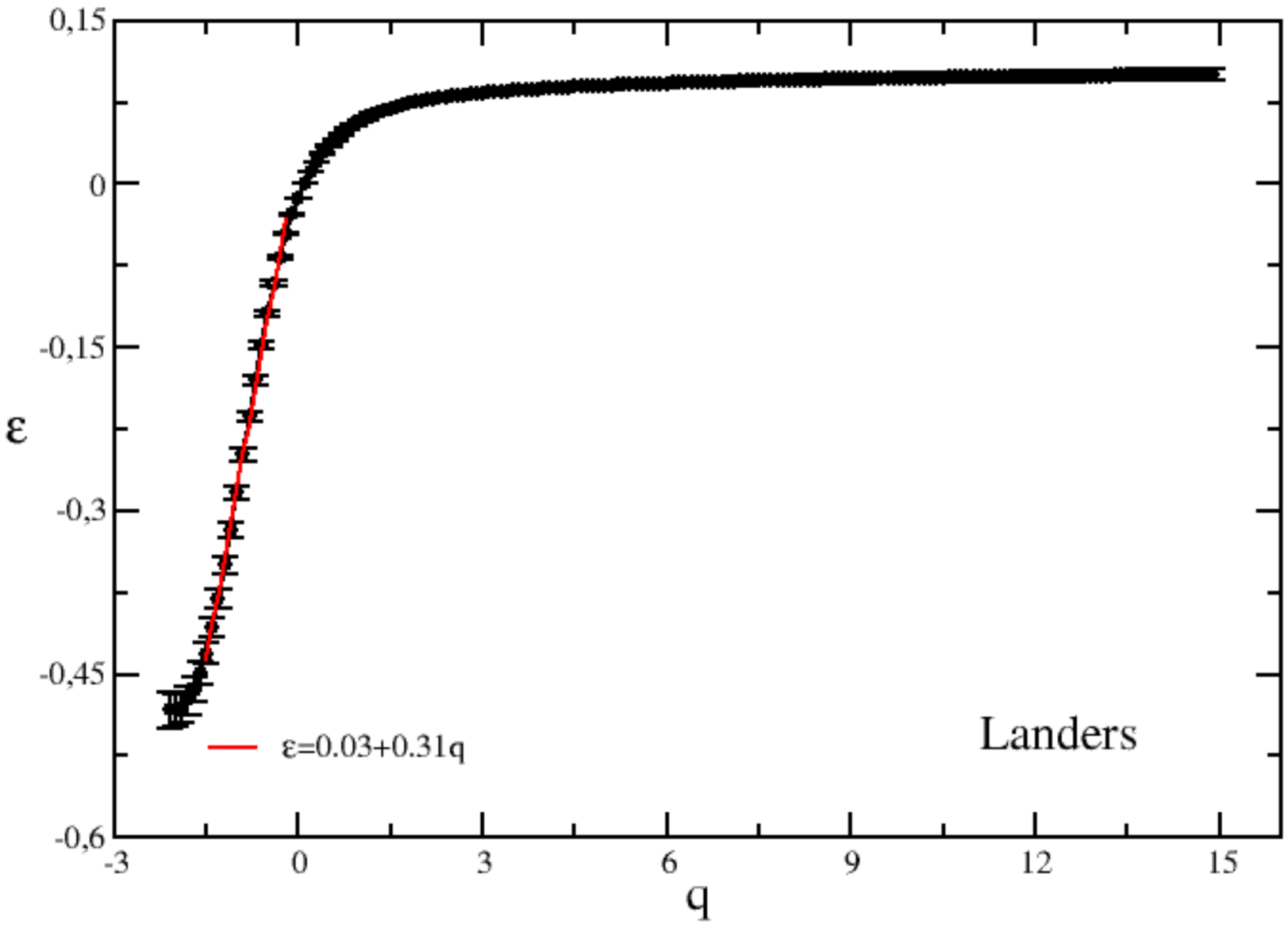

**Figure 3.** Scaling spectrum $\varepsilon(q)$ for the Landers earthquake sequence. The saturation at large positive $q$ indicates that the largest interevent distances exhibit a spatially bounded scaling regime.

The contrast between the nonlinear spectra of the earthquake swarms and the high-order saturation observed for Landers suggests that $\varepsilon(q)$ may provide a useful diagnostic for distinguishing different modes of clustered seismic migration.

The use of a common temporal fitting interval for all moment orders, together with the AICc-based selection of the polynomial degree for the positive branch, minimizes subjective choices in the estimation of $\varepsilon(q)$. The reported regression uncertainties therefore reflect the quality of the log-log fits, although they do not include the full statistical dependence among earthquake pairs.

## 4 Conclusions

In this work we investigated the multiscaling properties of earthquake migration in ten persistent seismic swarms identified within the Southern California relocated catalogue. Although the individual swarms differ in their spatial extent, temporal evolution, and detailed high-order spectra, they all display the same qualitative separation between an approximately linear negative-order branch and a nonlinear positive-order branch. The principal result is that the scaling spectrum $\varepsilon(q)$ systematically departs from the linear behavior expected for single-exponent diffusion. Negative-order moments are well described by an approximately linear spectrum, whereas positive-order moments exhibit a persistent nonlinear behavior, indicating that large and small interevent distances evolve through different effective scaling laws. This behavior is consistent with strong anomalous diffusion and with a heterogeneous migration process occurring on a complex spatial structure. A comparison with the Landers sequence further highlights the distinctive nature of earthquake swarms. Whereas Landers exhibits a high-order saturation of $\varepsilon(q)$, compatible with a spatially bounded diffusion process, the analyzed swarms maintain a nonlinear spectrum over the investigated range of moments. These results suggest that the multiscaling spectrum $\varepsilon(q)$ provides a quantitative fingerprint of earthquake migration and may represent a useful statistical tool for distinguishing fluid-driven seismic swarms from conventional tectonic earthquake sequences.

## Open Research Section

The source codes used for the identification of persistent seismic swarms and the multiscaling analysis presented in this study are openly available through the Zenodo repository at `https://doi.org/10.5281/zenodo.21297044`.

## Conflict of Interest declaration

The authors declare there are no conflicts of interest for this manuscript.

**Acknowledgments**
The author acknowledges support from the DyDéCo project and from the CPJ Chair at ENS de Lyon and CNRS.